\documentclass[preprint,preprintnumbers, prd, floatfix, superscriptaddress,nofootinbib] {revtex4-1}
\usepackage{epsfig}
\usepackage{subfigure}
\usepackage{dcolumn}
\usepackage{bm}
\usepackage[usenames ,dvipsnames]{xcolor}
\usepackage{slashed}
\usepackage{graphicx,color}
\usepackage{lineno}

\begin{document}
\title{Equivalent $SU(3)_f$ approaches
for two-body\\ anti-triplet charmed baryon decays}

\author{Y.~K. Hsiao}
\email{yukuohsiao@gmail.com}
\affiliation{School of Physics and Information Engineering, Shanxi Normal University, 
Taiyuan, 030031, China}

\author{Y.~L. Wang}
\email{1556233556@qq.com}
\affiliation{School of Physics and Information Engineering, Shanxi Normal University, 
Taiyuan, 030031, China}

\author{H.~J. Zhao}
\email{hjzhao@163.com}
\affiliation{School of Physics and Information Engineering, Shanxi Normal University, 
Taiyuan, 030031, China}

\date{\today}

\begin{abstract}
For the two-body ${\bf B}_c\to{\bf B}M$ decays,
where ${\bf B}_c$ denotes the anti-triplet charm baryon and ${\bf B}(M)$ the octet baryon (meson),
there exist two theoretical studies based on the $SU(3)$ flavor [$SU(3)_f$] symmetry.
One is the irreducible $SU(3)_f$ approach (IRA).
In the irreducible $SU(3)_f$ representation, the effective Hamiltonian related to 
the initial and final states forms the amplitudes for ${\bf B}_c\to{\bf B}M$.
The other is the topological-diagram approach (TDA), where
the $W$-boson emission and $W$-boson exchange topologies
are drawn and parameterized for the decays. 
As required by the group theoretical consideration, we present
the same number of the IRA and TDA amplitudes. 
We can hence relate the two kinds of the amplitudes, 
and demonstrate the equivalence of the two $SU(3)_f$ approaches. 
We find a specific $W$-boson exchange topology only contributing to $\Xi_c^0\to{\bf B}M$.
Denoted by $E_M$, it plays a key role in explaining 
${\cal B}(\Xi_c^0\to \Lambda^0 K_S^0,\Sigma^0 K_S^0,\Sigma^+ K^-)$. 
We consider that $\Lambda_c^+ \to n \pi^+$ and $\Lambda_c^+ \to p\pi^0$
proceed through the constructive and destructive interfering effects, respectively, 
which leads to ${\cal B}(\Lambda_c^+ \to n \pi^+)\gg{\cal B}(\Lambda_c^+ \to p\pi^0)$ 
in agreement with the data. With the exact and broken $SU(3)_f$ symmetries, 
we predict the branching fractions of ${\bf B}_c\to{\bf B}M$ to be tested by future measurements.
\end{abstract}

\maketitle
\section{introduction}
In recent years, more and more
charmed baryon decay channels have been reanalyzed and discovered~\cite{pdg,Belle:2021btl,Belle:2021gtf,
Belle:2021mvw,Belle:2021zsy,Belle:2021dgc,Belle:2021avh,BESIII:2022bkj}. 
The most observations come form ${\bf B}_c\to{\bf B}M$,
where ${\bf B}_c$ denotes the anti-triplet charmed baryon state, and ${\bf B}(M)$ the octet baryon (meson).
For example, Belle has newly reported the observation of 
$\Xi_c^0\to (\Lambda^0 K_S^0,\Sigma^0 K_S^0,\Sigma^+ K^-)$~\cite{Belle:2021avh},
and BESIII that of  $\Lambda_c^+ \to n \pi^+$~\cite{BESIII:2022bkj}. 
For investigation, theoretical attempts have been given, such as
the factorization~\cite{Gutsche:2018utw},
pole model~\cite{Cheng:2018hwl,Zou:2019kzq}, 
quark model~\cite{Korner:1992wi,Niu:2020gjw,Hsiao:2020gtc},  
current algebra~\cite{Groote:2021pxt}, and
final state interaction~\cite{Yu:2020vlt,Ke:2020uks},
whose calculations might be complicated. As the alternative approach,
the $SU(3)$ flavor $[SU(3)_f]$ symmetry can avoid the model calculation.

There exist two $SU(3)_f$ approaches for the ${\bf B}_c$ decays.
One is the irreducible $SU(3)_f$ approach~(IRA),
where the effective Hamiltonian (${\cal H}_{\text{eff}}$) for the $c$ quark decays can be 
presented as the irreducible $SU(3)_f$ expression. As a consequence, 
${\cal H}_{\text{eff}}$ connected to ${\bf B}_c$ and the final states
induces the $SU(3)_f$ invariant amplitudes~\cite{Savage:1989qr,
Savage:1991wu,Sharma:1996sc,Lu:2016ogy,Geng:2017mxn,Geng:2019awr,Geng:2019xbo,
Geng:2017esc,Geng:2018plk,Geng:2018upx,Hsiao:2019yur,Geng:2020zgr,Huang:2021aqu,
Jia:2019zxi,Wang:2017gxe,Wang:2019dls,Wang:2022kwe,Geng:2018bow}.
The other is the topological-diagram approach~(TDA)~\cite{Kohara:1991ug,Chau:1995gk,
Zhao:2018mov,Hsiao:2020iwc,Pan:2020qqo}, where
the $W$-boson emission $(W_{\text EM})$ and $W$-boson exchange $(W_{\text EX})$ effects
can be drawn and parameterized as the topological amplitudes.

It is reasonable to regard IRA and TDA as the equivalent approaches 
for the heavy hadron decays~\cite{Chau:1995gk,
Zeppenfeld:1980ex,He:2018php,He:2018joe,Hsiao:2020iwc,Wang:2020gmn}.
However, He and Wang first point out that the previous analyses 
using IRA and TDA could not consistently match~\cite{He:2018php}. 
In the two-body $D$ and $B$ decays, one seeks the overlooked TDA amplitudes 
to solve the mismatch problem~\cite{He:2018php,He:2018joe,Wang:2020gmn}.
In the two-body ${\bf B}_c\to{\bf B}M$ decays, the mismatch problem remains unsolved,
which is due to the inconclusive TDA amplitudes involved in the decays~\cite{Chau:1995gk,He:2018joe}.

The equivalence should be in accordance with the equal number of the IRA and TDA amplitudes.
For instance, one derives two $SU(3)_f$ amplitudes and two topological ones
for ${\bf B}_c\to{\bf B}^*M$~\cite{Hsiao:2020iwc}, where ${\bf B}^*$ denotes the decuplet baryon.
Without considering the singlet contributions to the formation of $\eta_1$,
there can be seven independent IRA amplitudes 
in the ${\bf B}_c\to{\bf B}M$ decays~\cite{Savage:1989qr,Geng:2017esc,He:2018joe},
whereas TDA leads to six, seven, eight and sixteen topological amplitudes
from Refs.~\cite{Zhao:2018mov}, \cite{Groote:2021pxt}, \cite{Kohara:1991ug,Chau:1995gk} 
and~\cite{He:2018joe}, respectively. Clearly, 
the unique unification of the two $SU(3)_f$ approaches is unavailable. 
Therefore, we propose to clarify how many independent topological amplitudes can actually exist,
and newly unify the IRA and TDA amplitudes.
We will perform the numerical analysis, in order to demonstrate that 
the new unification can accommodate the new data.

%
\begin{figure}
\includegraphics[width=1.8in]{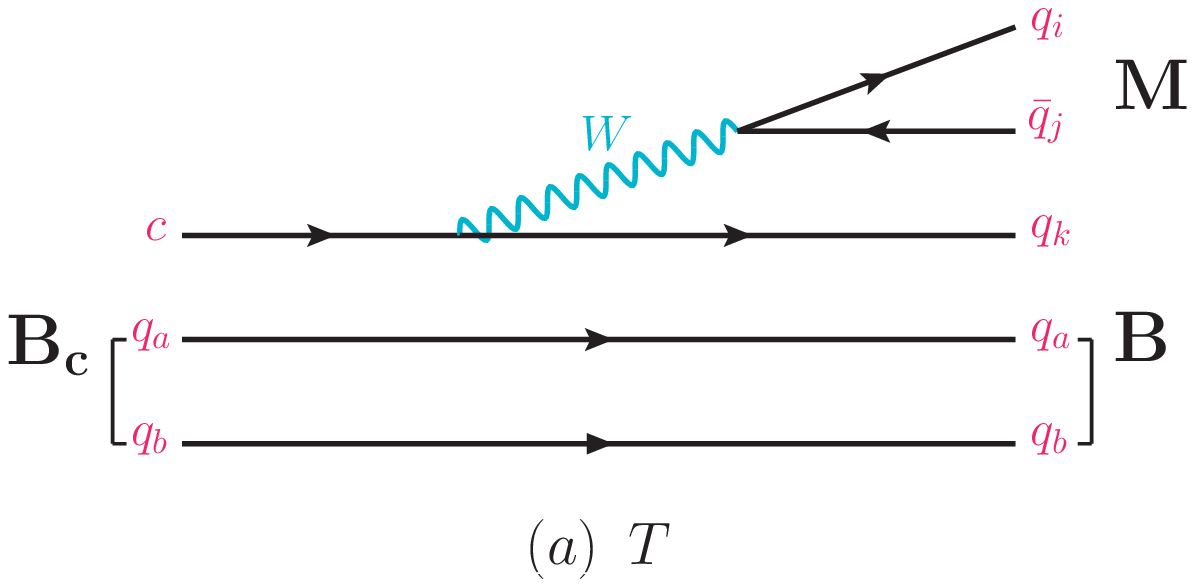}
\includegraphics[width=1.8in]{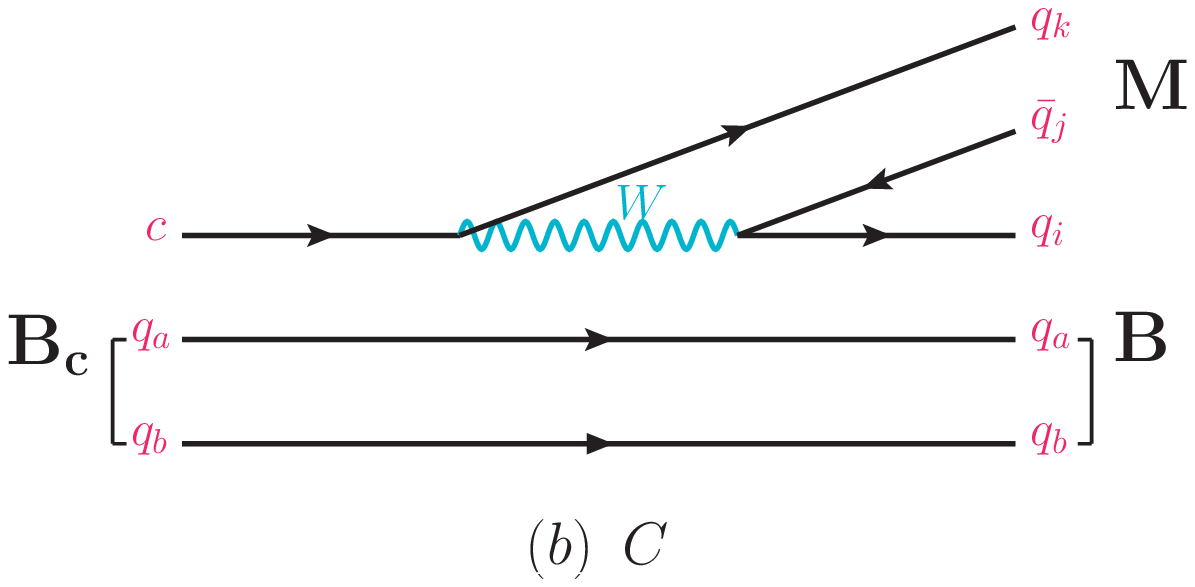}
\includegraphics[width=1.8in]{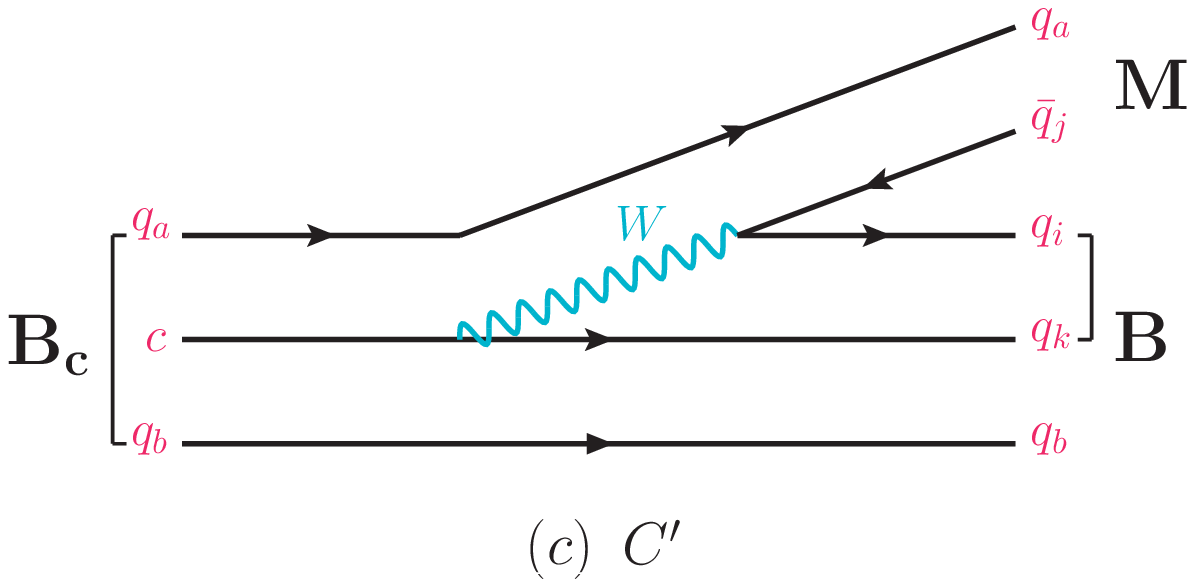}\\[2mm]
\includegraphics[width=1.8in]{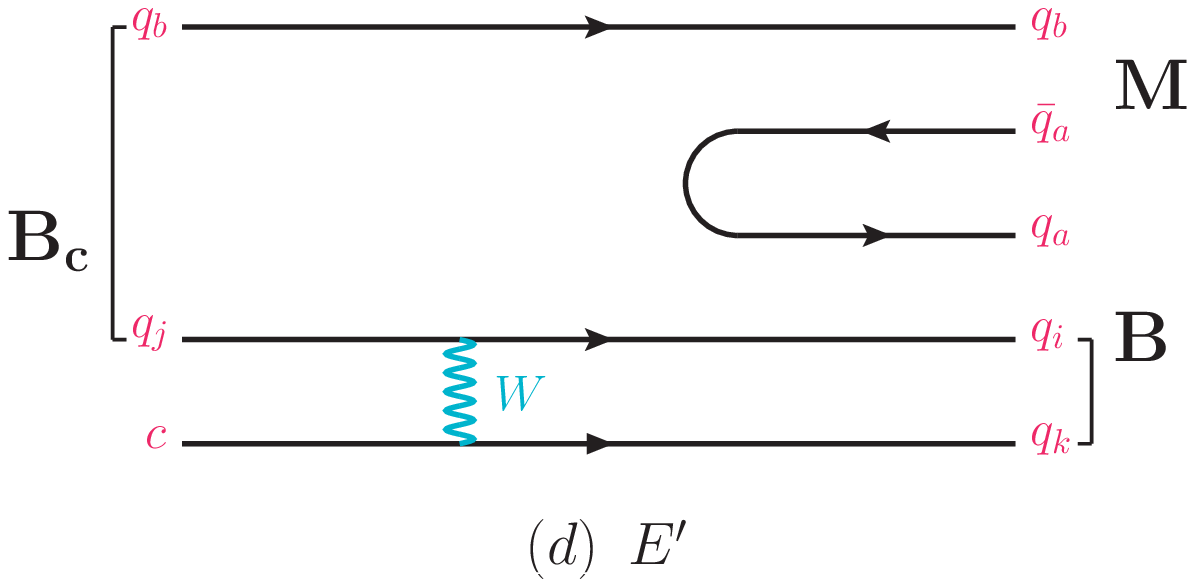}
\includegraphics[width=1.8in]{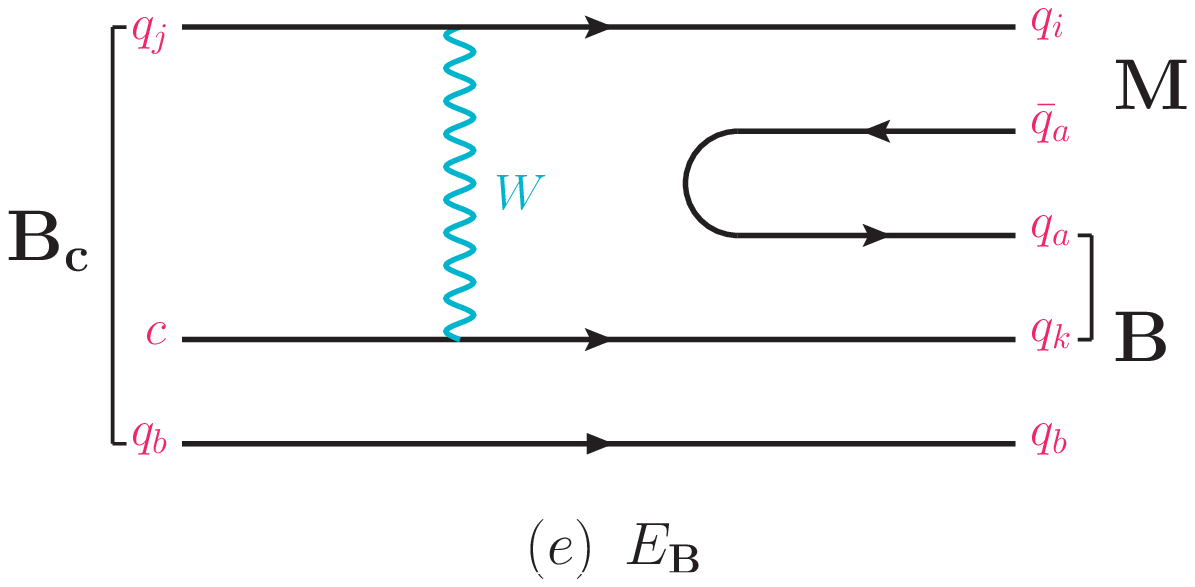}
\includegraphics[width=1.8in]{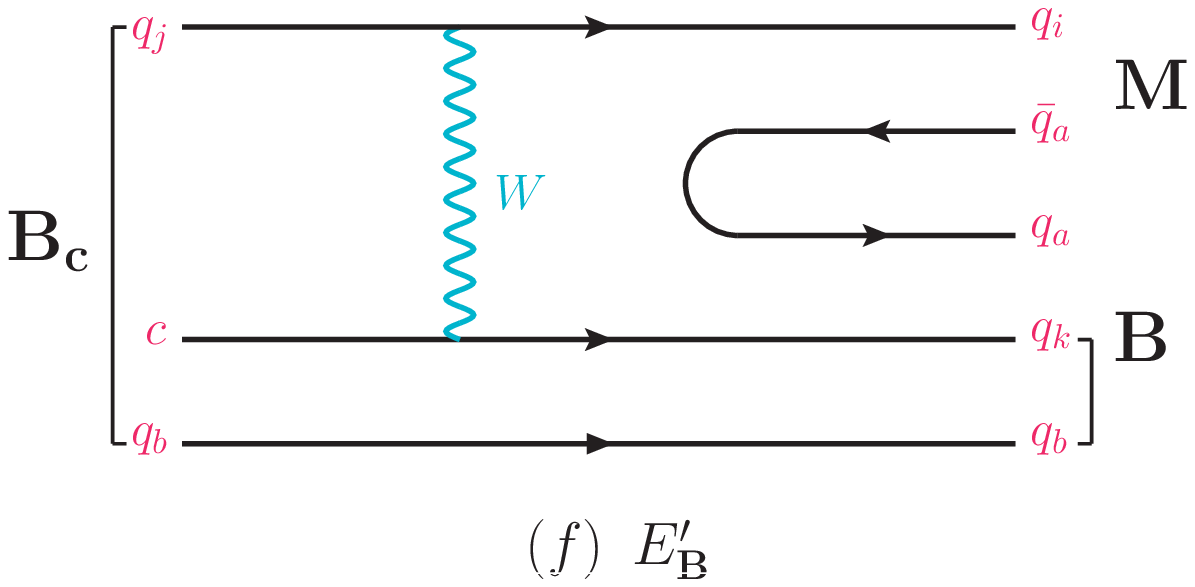}\\[2mm]
\includegraphics[width=1.8in]{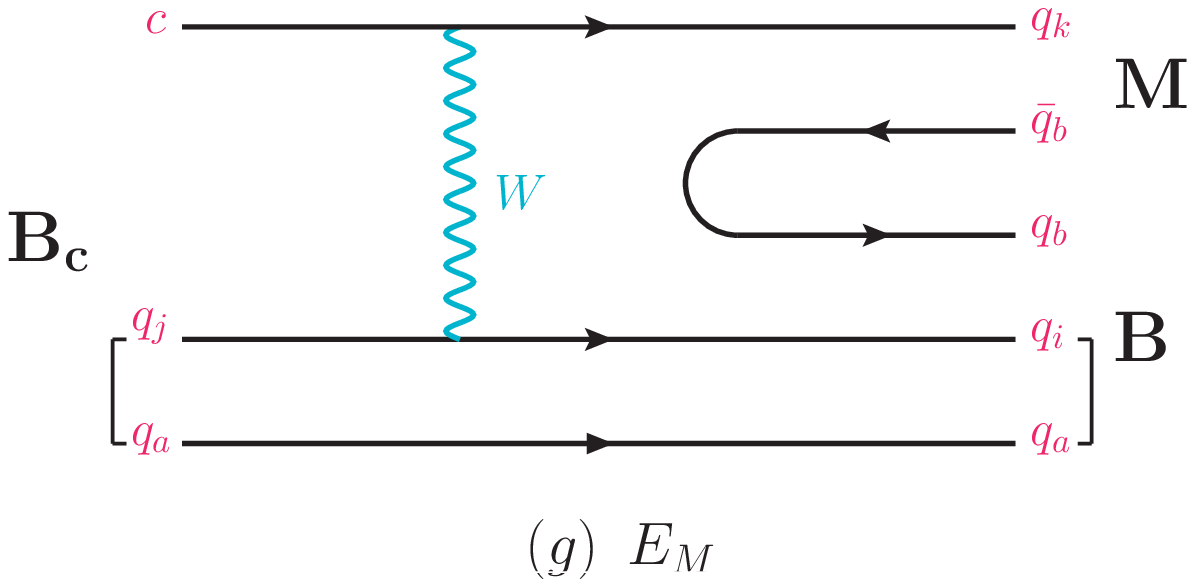}
\includegraphics[width=1.8in]{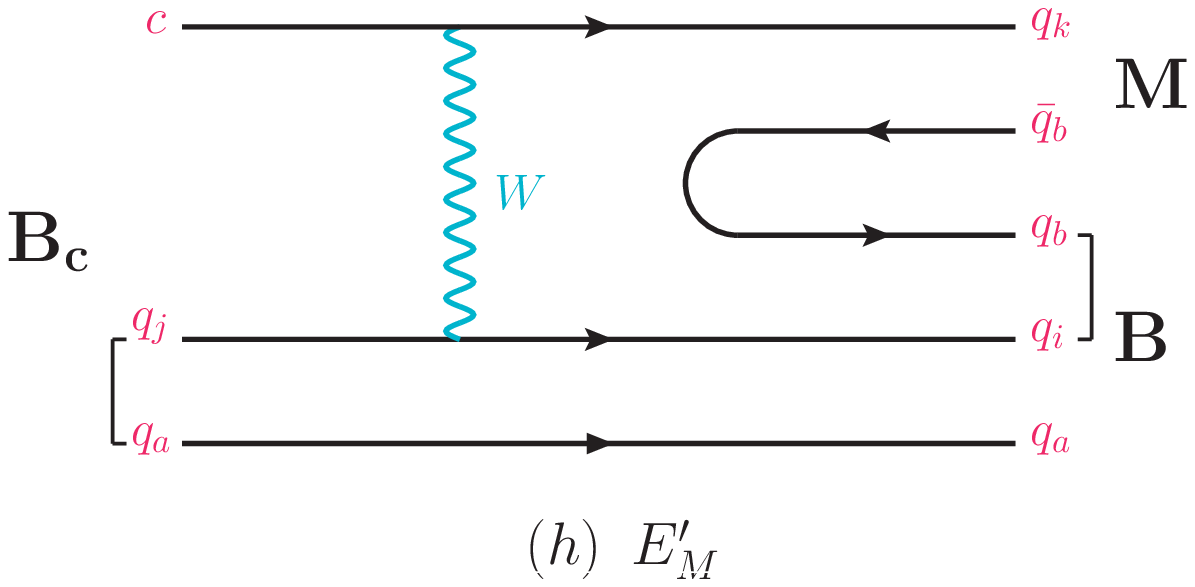}
\caption{Topological diagrams for the ${\bf B}_c\to {\bf B} M$ decays,
where the notations ``['' and ``]'' denote the anti-symmetric quark orderings
of the anti-triplet charmed and octet baryon states, respectively.}\label{fig1}
\end{figure}
%
%
\section{Formalism}
For the anti-triplet charmed baryon decays,
we present the effective Hamiltonian
for the $c\to u q \bar q'$ decays as~\cite{Buchalla:1995vs,Buras:1998raa}
\begin{eqnarray}\label{Heff}
{\cal H}_{c}\equiv{\cal H}_{eff}/(G_F/{\sqrt 2})&=&
\lambda_{q'q}[c_1(\bar u q' )(\bar q c)+c_2(\bar u_\beta q'_{ \alpha} )(\bar q_\alpha c_\beta)]\,,
\end{eqnarray}
where $G_F$ is the Fermi constant, and
$\lambda_{q'q}\equiv V_{uq'} V^*_{cq}$ denotes
the Cabibbo-Kobayashi-Maskawa (CKM) matrix elements.
Besides, $(\bar q_1 q_2)=\bar q_1\gamma_\mu(1-\gamma_5)q_2$
are quark currents, and  $(\alpha,\beta)$ represent the color indices.
As $q^{(\prime)}$ takes the values $d$ and $s$, the decays with
$\lambda_{q'q}=[V^*_{cs}V_{ud}$, $V^*_{cs(d)}V_{us(d)},V^*_{cd}V_{us}]
\simeq [1,s_c(-s_c),-s_c^2]$ are regarded as
the Cabibbo-allowed (CA), singly Cabibbo suppressed (SCS),
and doubly Cabibbo suppressed (DCS) processes, respectively,
where $\theta_c$ in $s_c\equiv \sin\theta_c\simeq 0.22$ is the Cabibbo angle.
Moreover, the Wilson coefficient $c_{1(2)}$ is a scale ($\mu$)-dependent number,
and we take $\mu=m_{c}$ in the $c$ decays.

By omitting Lorentz structure,
${\cal H}_c$ is seen as ${\cal H}_c\sim (\bar q_i q_j\bar q_k) c$,
where $q_i=(u,d,s)$ is a triplet~($3$) under the $SU(3)_f$ symmetry.
In IRA, ${\cal H}_c$ is decomposed as two irreducible $SU(3)_f$ forms,
$6_H$ and $\overline{15}_H$, whereas TDA only values
the flavor changes of $c\to q_i \bar q_j q_k$ in ${\cal H}_c$.
Thus, the effective Hamiltonian
can be rewritten as~\cite{Savage:1989qr,Savage:1991wu,Geng:2017esc,Pan:2020qqo,Hsiao:2020iwc}
\begin{eqnarray}\label{Heff2}
{\cal H}_{\text{IRA}}=c_- { \epsilon^{ijl} \over 2}H(6)_{lk}+c_+H(\overline{15})_{k}^{ij}\,,\;
{\cal H}_{\text{TDA}}=H_j^{ki}\,,
\end{eqnarray}
where $c_\mp=(c_1\mp c_2)$, and
the non-zero entries of $H(6)_{lk}$, $H(\overline{15})_k^{ij}$ and $H_j^{ki}$
are given by~\cite{Savage:1989qr,Hsiao:2020iwc}
\begin{eqnarray}
&&
H_{22}(6)=2\,,\;H_2^{13}(\overline{15})=H_2^{31}(\overline{15})=1\,,\;\nonumber\\
&&
H_{23,32}(6)=-2s_c\,,H_2^{12,21}(\overline{15})=-H_3^{13,31}(\overline{15})=s_c\,,\nonumber\\
&&
H_{33}(6)=2s_c^2\,,\;H_3^{12,21}(\overline{15})=-s_c^2\,,\;\nonumber\\
&&
H^{31}_2=1\,,\; H^{21}_2=-s_c\,,\;
H^{31}_3=s_c\,,\; H^{21}_3=s_c^2\,.\;
\end{eqnarray}
As the final states, the octet baryon and meson ($8_{\bf B}$ and $8_M$)
have components
\begin{eqnarray}
{\bf B}^i_j&:&(n,\;p,\;,\Sigma^{\pm,0}, \Xi^{-,0},\;\Lambda)\,,\nonumber\\
M^i_j&:&(\pi^{\pm,0},\;K^\pm,\;K^0,\;\bar K^0,\;\eta)\,,
\end{eqnarray}
where the octet baryon can also be written as ${\bf B}_{ijk} = \epsilon_{ijl} {\bf B}^l_k$.
Moreover, the $\eta$ state mixes $\eta_q=\sqrt{1/2}(u\bar u+d\bar d)$ and $\eta_s=s\bar s$
as $\eta=\eta_q\cos\phi-\eta_s\sin\phi$, where
$\phi=(39.3\pm1.0)^\circ$ is the mixing angle~\cite{FKS}.
For the anti-triplet charmed baryon states ($\bar 3_c$),
$\Xi_c^0$, $\Xi_c^+$ and $\Lambda_c^+$ consist of
$(ds-sd)c$, $(su-us)c$ and $(ud-du)c$, respectively,
and we present them in the two forms,
\begin{eqnarray}\label{3c}
{\bf B}_c({\bf B}_{c\;i})&=&(\Xi_c^0,-\Xi_c^+,\Lambda_c^+)\,,\nonumber\\
{\bf B}_c({{\bf B}_c}^{ij})&=&\left(\begin{array}{ccc}
0& \Lambda_c^+ & \Xi_c^+\\
-\Lambda_c^+&0&\Xi_c^0 \\
-\Xi_c^+&-\Xi_c^0&0
\end{array}\right)\,.
\end{eqnarray}
We can hence construct the amplitudes of ${\bf B}_c\to{\bf B}M$ in IRA and TDA,
written as~\cite{Kohara:1991ug,He:2018joe,Zhao:2018mov,Pan:2020qqo,Hsiao:2020iwc}
\begin{eqnarray}\label{amp1}
{\cal M}_{\text{IRA}}&=&
{\cal M}_{6}+{\cal M}_{\overline{15}}\,,\nonumber\\
{\cal M}_{6}&=&
a_1 H_{ij}(6)T^{ik}{\bf B}_k^l M_l^j+
a_2 H_{ij}(6)T^{ik}M_k^l {\bf B}_l^j+
a_3 H_{ij}(6){\bf B}_k^i M_l^j T^{kl}\,,\nonumber\\
{\cal M}_{\overline{15}}
&=&
a_4H_{k}^{li}(\overline{15}){\bf B}_{c\,j} M_i^j {\bf B}^l_k+
a_5{\bf B}^i_j M^l_i H(\overline{15})^{jk}_l {\bf B}_{c\,k}\nonumber\\
&+&
a_6{\bf B}^k_l M^i_j H(\overline{15})^{jl}_i {\bf B}_{c\,k}
+a_7{\bf B}^l_i M^i_j H(\overline{15})^{jk}_l {\bf B}_{c\,k}\,,\nonumber\\
{\cal M}_{\text{TDA}}
&=&
T{{\bf B}_c}^{ab} H^{ki}_j{\bf B}_{abk}M_j^i
+C {{\bf B}_c}^{ab} H^{ki}_j{\bf B}_{abi}M_j^k
+C^\prime{{\bf B}_c}^{ab} H^{ki}_j{\bf B}_{ikb}M_j^a\,\nonumber\\
&+&
E_{\bf B} {{\bf B}_c}^{jb} H^{ki}_j{\bf B}_{kab}M_a^i+
E_{\bf B}^{\prime}{{\bf B}_c}^{jb} H^{ki}_j{\bf B}_{kba}M_a^i\,\nonumber\\
&+&
E_{M}{{\bf B}_c}^{jb} H^{ki}_j{\bf B}_{iba}M_a^k+
E_{M}^{\prime} {{\bf B}_c}^{jb} H^{ki}_j{\bf B}_{iab}M_a^k+
E^{\prime}{{\bf B}_c}^{jb} H^{ki}_j{\bf B}_{ika}M_a^b\,,
\end{eqnarray}
where $T^{ij} \equiv{\bf B}_{c\,k}\epsilon^{ijk}$,
$a_i$ ($i=1,2, ...,7$) are the $SU(3)_f$ invariant amplitudes,
and $(T,C^{(\prime)}$, $E_{{\bf B}(M)}^{(\prime)},E')$ the topological ones.

For ${\cal M}_{\text{IRA}}$, $(c_-,c_+)$ in ${\cal H}_c$ are absorbed into $a_i$ ($i=1,2,...,7$).
Since one obtains $(c_-,c_+)=(1.65,0.79)$ as a result of QCD corrections~\cite{Li:2012cfa},
$(a_1,a_2,a_3)$ with $c_-$ [$(a_4,a_5,a_6,a_7)$ with $c_+$]
are regarded as QCD-(dis)favored parameters. Under the group theoretical consideration,
we present $3_c\times 6_H \times 8_{\bf B}\times 8_M$ and $3_c\times \overline{15}_H \times 8_{\bf B}\times 8_M$
for ${\bf B}_c\to{\bf B}M$~\cite{Pan:2020qqo}, which lead to 3 and 4 $SU(3)_f$ singlets 
to be in accordance with $a_{1,2,3}$ and $a_{4,5,6,7}$, respectively. 
Hence, there can be seven independent IRA amplitudes 
as those in~\cite{Savage:1989qr,Geng:2017esc,He:2018joe}.

For ${\cal M}_{\text{TDA}}$, the topological amplitudes correspond to 
the $W$-boson emission $(W_{\text EM})$ and $W$-boson exchange $(W_{\text EX})$ diagrams
in Fig.~\ref{fig1}(a-c) and Fig.~\ref{fig1}(d-h), respectively. More specifically, 
$T$ and $C^{(\prime)}$ represent the external and internal $W_{EM}$ processes
in Fig.~\ref{fig1}a and Fig.~\ref{fig1}b(c), respectively. Compared to $T$ and $C^{(\prime)}$,
the $W_{EX}$ process ($E$) needs an additional gluon to connect a quark pair in ${\bf B}M$.
Particularly, $E_{{\bf B}(M)}$ with the subscript ${\bf B}(M)$
stands for the $W_{EX}$ topology, where the baryon (meson) receives the quark $q_k$
from the $c\to q_k$ transition~$[$see Fig.~\ref{fig1}e(g)$]$. Moreover, 
$E_{{\bf B}(M)}'$ presents the same topology as $E_{{\bf B}(M)}$
except that their baryon states have different quark orderings,
that is, ${\bf B}\sim (q_a-q_k)q_b\,[(q_i-q_a)q_b]$ for $E_{{\bf B}(M)}$ and 
${\bf B}\sim (q_k-q_b)q_a\,[(q_b-q_i)q_a]$ for $E_{{\bf B}(M)}'$. 
By contrast, $E'$ parameterizes the $W_{EX}$ process in Fig.~\ref{fig1}d,
where the meson receives no quark to relate to the $W$-boson.

Based on the $SU(3)_f$ symmetry, TDA and IRA should have seven independent amplitudes.
Nonetheless, we derive eight TDA amplitudes. 
Since we find the appearance of $(E'-E_{\bf B}^\prime)$ in the ${\cal M}_{\text{TDA}}$ expansion,
either $E'$ or $E_{\bf B}^\prime$ can be redundant. 
We thus reduce the TDA amplitudes to seven 
by choosing to work with the convention of $E_{\bf B}^\prime=0$.
Using ${\cal M}_{\text{IRA}}={\cal M}_{\text{TDA}}$, 
we find the relations as
\begin{eqnarray}\label{relation1}
&&(T,C,C')=(a_1+\frac{a_4+a_6}{2},-a_1+\frac{a_4+a_6}{2},-2a_1+2a_3)\,,\nonumber\\
&&(E_{\bf B},E_M,E')=(a_4,-2a_4-2a_7,-2a_2+a_4+a_7)\,,
\end{eqnarray}
and $E_M^\prime=E_{\bf B}$. We also obtain
\begin{eqnarray}\label{relation2}
&&(a_1,a_2,a_3)=(\frac{T-C}{2},-\frac{E_M+2E'}{4},\frac{T-C+C'}{2})\,,\nonumber\\
&&(a_4,a_6,a_7)=(E_{\bf B},T+C-E_{\bf B},-E_{\bf B}-{E_M\over 2})\,,
\end{eqnarray}
and $a_5=0$, such that we ``topologize'' the $SU(3)_f$ invariant amplitudes.

The $W_{EX}$ diagrams require $g\to q\bar q$ added to ${\bf B}M$, where $q\bar q$ can be
$u\bar u$, $d\bar d$, or $s\bar s$. Due to $m_s\gg m_{u,d}$,
it is possible that the exchange topologies with $g\to s\bar s$ can cause 
the sizeable $SU(3)_f$ symmetry breaking. 
Therefore, 
we denote $E_{{\bf B},M}$, $E'_M$ and $E'$ with $g\to s\bar s$
by $E_{{\bf B},M}^{(s)}$, $E_M^{\prime (s)}$ and $E^{\prime (s)}$, respectively,
for the ${\bf B}_c\to{\bf B}M$ decays. Note that
one has included the $SU(3)_f$ breaking effect via the $W_{EX}$ diagrams,
and demonstrated that it can be applicable to
$D\to PP(PV)$~\cite{Cheng:2019ggx} and ${\bf B}_c\to{\bf B}^*M$~\cite{Hsiao:2020iwc}.

In terms of the equations given by~\cite{pdg}
\begin{eqnarray}\label{p_space}
&&{\cal B}({\bf B}_c\to{\bf B}M)=
\frac{G_F^2|\vec{p}_{{\bf B}}|\tau_{{\bf B}_c}}{16\pi m_{{\bf B}_c}^2 }
|{\cal M}({\bf B}_c\to{\bf B} M)|^2\,,\nonumber\\
&&|\vec{p}_{{\bf B}}|=\frac{
\sqrt{(m_{{\bf B}_c}^2-m_+^2)(m_{{\bf B}_c}^2-m_-^2)}}{2 m_{{\bf B}_c}}\,,
\end{eqnarray}
we can compute the branching fraction to be used in the numerical analysis,
where $m_\pm=m_{{\bf B}}\pm m_M$, and
$\tau_{{\bf B}_c}$ stands for the ${\bf B}_c$ lifetime,
${\cal M}({\bf B}_c\to{\bf B} M)$ is from the full expansion of 
${\cal M}_{\text{IRA}}$ and ${\cal M}_{\text{TDA}}$  
in Tables~\ref{tab1}, \ref{tab2} and \ref{tab3}.

\begin{table}[t!]
\caption{Cabibbo-allowed (CA) amplitudes
in the expansions of ${\cal M}_{\text{TDA}}$ and ${\cal M}_{\text{IRA}}$, and
$(s\phi,c\phi)\equiv (\sin\phi,\cos\phi)$.}\label{tab1}
\tiny
\begin{tabular}{|l|l|l|}
\hline
Decay mode
&$\;\;\;\;\;\;\;\;\;\;\;\;\;\;\;\;\;\;\;\;\;\;\;\;$ $M_{\text{TDA}}$
&$\;\;\;\;\;\;\;\;\;\;\;\;\;\;\;\;\;\;\;\;\;\;\;\;$ $ M_{\text{IRA}}$ \\
\hline\hline
$\Lambda_c^+ \to \Lambda^0 \pi^+$
&$-\frac{1}{\sqrt{6}}(4T+C^\prime-E_{{\bf B}}-E^\prime)$
& $-\frac{\sqrt{6}}{3}(a_1+a_2+a_3-\frac{a_{5}- 2a_{6}+a_7}{2})$
\\
$\Lambda_c^+ \to \Sigma^0 \pi^+$
&$\frac{1}{\sqrt{2}}(C^\prime +E_{{\bf B}}-E^\prime)$
&$-\sqrt{2}(a_1-a_2-a_3-\frac{a_5-a_7}{2})$
\\
$\Lambda_c^+ \to \Sigma^+ \pi^0$
&$-\frac{1}{\sqrt 2}(C^\prime +E_{{\bf B}}-E^\prime)$
& $\sqrt{2}(a_{1}-a_{2}-a_{3}-\frac{a_{5}-a_{7}}{2})$
\\
$\Lambda_c^+ \to \Xi^0 K^+$
&$E^{\prime(s)}$
& $-2(a_2-\frac{a_{4}+ a_{7}}{2})$
\\
$\Lambda_c^+ \to p \bar K^0$
&$2C-E_M^\prime$
& $-2(a_1-\frac{a_{5}+a_{6}}{2})$
\\
$\Lambda_c^+ \to \Sigma^+ \eta$
&$[\frac{1}{\sqrt 2}(C^\prime-E_{{\bf B}}+E^\prime)c\phi- E_M^{\prime(s)} s\phi]$
& $\sqrt{2}c\phi(-a_1-a_2+a_3+\frac{a_5+a_7}{2})-s\phi a_4$
\\
%
%
\hline
$\Xi_c^+ \to \Sigma^+ \bar{K}^0$
&$-2C+C^\prime$
&$ 2(a_{3}-\frac{a_{4} + a_{6}}{2})$
\\
$\Xi_c^+ \to \Xi^0 \pi^+$
&$-2T-C^\prime$
& $2(a_{3}+\frac{a_{4} + a_{6}}{2})$
\\
\hline
$\Xi_c^0 \to \Lambda^0 \bar{K}^0$
&$\frac{1}{\sqrt 6}(2C+C^\prime -E_{M}-2E_M^\prime-E^\prime)
$&$-\frac{\sqrt{6}}{3}(2a_1-a_2-a_3$
$+\frac{2a_5-a_6-a_7}{2})
$\\
$\Xi_c^0 \to  \Sigma^0\bar{K}^0$
&$\frac{1}{\sqrt 2}(2C-C^\prime +E_{M}+E^\prime)$
&$-\sqrt{2}(a_{2}+a_{3}-\frac{a_{6}-a_{7}}{2})$
\\
$\Xi_c^0 \to \Sigma^+ K^-$
&$-E_{M}-E^\prime$
& $ 2(a_{2}+\frac{a_{4} + a_{7}}{2})$
 \\
$\Xi_c^0 \to \Xi^0 \pi^0$
&$\frac{1}{\sqrt 2}(E_{{\bf B}}+C^\prime)$
& $ -\sqrt{2}(a_{1}-a_{3}$
$-\frac{a_{4}-a_{5}}{2})
$ \\
$\Xi_c^0 \to \Xi^-\pi^+$
&$2T-E_{{\bf B}}
$& $ 2(a_{1}+\frac{a_{5} + a_{6}}{2})
$  \\
$\Xi_c^0 \to \Xi^0 \eta $
&$[\frac{1}{\sqrt{2}}(E_{{\bf B}}-C^\prime)c\phi+(E_{M}^{(s)}+E_M^{\prime(s)}+E^{\prime(s)})s\phi]$
& $ \sqrt{2}c\phi(a_1- a_3 + \frac{a_4+a_5}{2} )$
$-2s\phi(a_2 + \frac{a_7}{2} ) $ \\
%
\hline
\end{tabular}
\end{table}
%
%
\begin{table}[t!]
\caption{Singly Cabibbo-suppressed (SCS) amplitudes
in the expansions of ${\cal M}_{\text{TDA}}$ and ${\cal M}_{\text{IRA}}$.}\label{tab2}
\tiny
\begin{tabular}{|l|l|l|}
\hline
Decay mode
&$\;\;\;\;\;\;\;\;\;\;\;\;\;\;\;\;\;\;\;\;\;\;\;\;\;\;\;\;\;\;\;\;\;\;$ $M_{\text{TDA}}$
&$\;\;\;\;\;\;\;\;\;\;\;\;\;\;\;\;\;\;\;\;\;\;\;\;\;\;\;\;\;\;\;\;\;\;$ $ M_{\text{IRA}}$ \\
\hline\hline
$\Lambda_c^+ \to \Lambda^0 K^+$
&$[- \frac{1}{\sqrt{6}}(-E^{(s)}_{{\bf B}}+E^{\prime(s)}+4T+C^\prime)]s_c$
&$-\frac{\sqrt{6}}{3}(a_1-2a_2+a_3+\frac{3a_4-a_5+2a_6+2a_7}{2})s_c$
\\
$\Lambda_c^+ \to \Sigma^0 K^+$
&$[\frac{1}{\sqrt{2}}(E_{\bf B}^{(s)}+C^\prime)]s_c$
&$ -[\sqrt 2(a_{1}-a_3-\frac{a_{4} + a_{5}}{2})]s_c$
\\
$\Lambda_c^+ \to \Sigma^+ K^0$
&$(-E_M^{\prime\,(s)}+C^\prime)s_c$
&$ -2(a_{1}-a_3+\frac{a_{4} - a_{5}}{2})s_c$
\\
$\Lambda_c^+ \to n \pi^+$
&$(-2T-C^\prime+E^\prime)s_c$
&$-2(a_{2}+a_3+\frac{a_6 - a_7}{2})s_c$
\\
$\Lambda_c^+ \to p \pi^0$
&$[\frac{1}{\sqrt 2}(2C-C^\prime-E_{{\bf B}}-E_M^\prime+E^\prime)]s_c $
&$-[\sqrt 2(a_{2}+a_3-\frac{a_{6}+a_{7}}{2})]s_c$
\\
$\Lambda_c^+ \to p \eta$
&$[-\frac{1}{\sqrt 2}(2C-C^\prime-E_M^\prime+E_{\bf B}-E^\prime)c\phi-2Cs\phi]s_c$
&$-[\sqrt{2}c\phi(a_2-a_3+\frac{a_6-a_7}{2})-2s\phi(a_1-\frac{a_4+a_5+a_6}{2})]s_c $ \\
%
\hline
$\Xi_c^+ \to \Lambda^0 \pi^+$
&$\frac{1}{\sqrt{6}}(T+C^\prime+E_{{\bf B}}+E^\prime)s_c$
& $-\frac{\sqrt{6}}{3}(a_1+a_2-2a_3-\frac{3a_4+a_5+a_6+a_7}{2})s_c$
\\
$\Xi_c^+ \to \Sigma^0  \pi^+$
&$\frac{1}{\sqrt{2}}(-2T+E_{{\bf B}}-E^\prime)s_c$
&$-\sqrt 2(a_{1}-a_{2}+\frac{a_{4}-a_{5}+a_{6}+a_{7}}{2})s_c$
\\
$\Xi_c^+ \to \Sigma^+ \pi^0$
&$\frac{1}{\sqrt 2}(-2C-E_{{\bf B}}+E^\prime)s_c$
&$\sqrt 2(a_{1}-a_{2}-\frac{a_{4}+a_{5}+a_{6}-a_{7}}{2})s_c$
\\
$\Xi_c^+ \to \Xi^0 K^+$
&$-(2T+C^\prime-E^{\prime(s)})s_c$
& $-2(a_2+a_{3}+\frac{a_{6} - a_{7}}{2})s_c$
\\
$\Xi_c^+ \to p \bar{K}^0$
&$(C^\prime-E_M^\prime)s_c$
& $-2(a_1-a_{3}+\frac{a_{4} - a_{5}}{2})s_c$
\\
$\Xi_c^+ \to \Sigma^+ \eta$
&$[\frac{1}{\sqrt{2}}(2C-E_{{\bf B}}+E^\prime)c\phi+(2C-C^\prime-E_M^{\prime(s)})s\phi]s_c$
&$-[ \sqrt{2}c\phi(a_1+a_2-\frac{a_4+a_5+a_6+a_7}{2})+2s\phi(a_3-\frac{a_6}{2})]s_c$
\\
\hline
$\Xi_c^0 \to \Sigma^+ \pi^-$
&($E_{M}+E^\prime)s_c$
& $-2(a_{2}+\frac{a_{4} + a_{7}}{2})s_c$
\\
$\Xi_c^0 \to \Sigma^- \pi^+$
&($-2T+E_{{\bf B}})s_c$
&$-2(a_1+\frac{a_{5}+a_{6}}{2})s_c$
\\
$\Xi_c^0 \to \Lambda^0 \pi^0$
&$[-\frac{1}{2\sqrt 3}(-2C+2C^\prime+3E_{{\bf B}}+2E_M^\prime+E_{M}+E^\prime)]s_c$
&$\sqrt{\frac{1}{3}}(a_1+a_2-2a_3-\frac{3a_4-a_5-a_6-a_7}{2})s_c$
\\
$\Xi_c^0 \to \Sigma^0 \pi^0$
&$\frac{1}{2}(2C+E_{M}+E_B+E^\prime)s_c$
&$-(a_1+a_2-\frac{a_4-a_5+a_6-a_7}{2})s_c$
\\
$\Xi_c^0 \to \Xi^-K^+$
&$(-E_{{\bf B}}^{(s)}+2T)s_c$
& $ 2(a_{1}+\frac{a_{5} + a_{6}}{2})s_c$
\\
$\Xi_c^0 \to \Xi^0 K^0$
&$(E_{M}^{(s)}+E_M^{\prime(s)}-C^\prime+E^{\prime(s)})s_c$
& $2(a_{1}-a_2-a_{3}$
$+\frac{a_{5}-a_{7}}{2})s_c$
\\
$\Xi_c^0 \to p K^- $
&$(-E^\prime-E_{M} )s_c$
&$2(a_{2}+\frac{a_{4} + a_{7}}{2})s_c$
\\
$\Xi_c^0 \to n \bar{K}^0$
&$(C^\prime-E^\prime-E_{M}-E_M^\prime)s_c$
& $-2(a_{1}-a_{2}-a_{3}+\frac{a_{5}-a_{7}}{2})s_c$
\\
$\Xi_c^0 \to \Lambda^0 \eta$
&$-[\frac{1}{2\sqrt 3}(-2C+2C^\prime+E_{M}-3E_{\bf B}+E_M^\prime+E^\prime)c\phi$
&$ -[\frac{\sqrt{3}c\phi}{3}(a_1+a_2-2a_3+\frac{3a_4+a_5+a_6+a_7}{2}) $\\
&$+\frac{1}{\sqrt{6}}(-2E^{\prime(s)}-2C-C^\prime-2E_{M}^{(s)}-E_{M}^{\prime(s)})s\phi]s_c$
&$-\frac{\sqrt{6}s\phi}{3}(2a_1+2a_2-a_3+\frac{2a_5-a_6+2a_7}{2})]s_c$
\\
%
$\Xi_c^0 \to \Sigma^0 \eta$
&[$\frac{1}{2}(-2C+E_{M}+E_{B}-E^\prime)c\phi-\frac{1}{\sqrt{2}}(2C-C^\prime-E_M^{\prime(s)})s\phi]s_c$
& $-[-c\phi(a_1+a_2+\frac{a_4+a_5-a_6+a_7}{2})-\sqrt{2}s\phi(a_3-\frac{a_6}{2})]s_c$
\\
%
%
\hline
\end{tabular}
\end{table}
%
\section{Numerical Analysis}
In the numerical analysis, we adopt the CKM matrix elements and
$(m_{{\bf B}_{(c)}},\tau_{{\bf B}_c})$ from PDG~\cite{pdg},
where
\begin{eqnarray}\label{B1}
&&(V_{cs},V_{ud},V_{us},V_{cd})=(1-\lambda^2/2,1-\lambda^2/2,\lambda,-\lambda)\,,
\end{eqnarray}
with $\lambda=s_c=0.22453\pm 0.00044$ in the Wolfenstein parameterization.
Making use of
\begin{eqnarray}\label{chi2}
\chi^2=\Sigma_i[({\cal B}_{th}^i-{\cal B}_{ex}^i)/\sigma_{ex}^{i}]^2\,,
\end{eqnarray}
we perform a minimum $\chi^2$-fit. As the theoretical input,
${\cal B}_{th}$ is calculated with the equations in Eq.~(\ref{p_space}) and
${\cal M}({\bf B}_c\to{\bf B}M)$ in Tables~\ref{tab1} and \ref{tab2}.
As the experimental input, ${\cal B}_{ex}$ can be found in Tables~\ref{BF_CA} and \ref{BF_SCS},
along with $\sigma_{ex}$ the experimental error.
Until very recently, only the upper limit of ${\cal B}_{ex}(\Lambda_c^+\to p\pi^0)>0.8\times 10^{-4}$ 
has been reported by Belle in Ref.~\cite{Belle:2021mvw}, from which 
the likelihood distribution in Fig.~7 as a function of ${\cal B}(\Lambda_c^+\to p\pi^+)$
can be used to estimate that ${\cal B}_{ex}(\Lambda_c^+\to p\pi^+)=(0.3\pm 0.3)\times 10^{-4}$.

We perform the global fit in the two scenarios.
In the first scenario (S1), we exactly preserve the $SU(3)_f$ symmetry,
such that $E_{{\bf B},M}^s=E_{{\bf B},M}$, $E_M^{\prime s}=E_M^{\prime}$ 
and $E^{\prime s}=E^{\prime}$.
The topological amplitudes as complex numbers can be written as
\begin{eqnarray}\label{para}
|T|, |C|e^{i\delta_C},|C'|e^{i\delta_{C'}},
|E_{\bf B}|e^{i\delta_{E_{\bf B}}}, |E_M| e^{i\delta_{E_M}}, |E'| e^{i\delta_{E'}}\,,
\end{eqnarray}
where $T$ has been set as a relatively real number, 
and $E'_M=E_{\bf B}$ has been implied in Eq.~(\ref{relation1}).

In the second scenario~(S2), we test the $SU(3)_f$ symmetry breaking,
which is indicated by the ratio of 
${\cal B}(\Xi_c^0\to \Xi^- K^+)$ to ${\cal B}(\Xi_c^0\to \Xi^- \pi^+)$,
\begin{eqnarray}\label{RXic}
{\cal R}(\Xi_c^0)&\equiv&
\frac{{\cal B}(\Xi_c^0\to \Xi^- K^+)}{{\cal B}(\Xi_c^0\to \Xi^- \pi^+)}
=s_c^2\frac{(2T-E_{{\bf B}}^{(s)})^2}{(2T-E_{{\bf B}})^2}\,.
\end{eqnarray}
With $E_{\bf B}^s=E_{\bf B}$, we obtain ${\cal R}(\Xi_c^0)=s_c^2\simeq 0.05$
away from the data of $0.03\pm 0.01$ by two standard deviation.
Note that the IRA amplitudes would cause the same deviation,
when one investigates $\Xi_c^0\to \Xi^- K^+$ and $\Xi_c^0\to \Xi^- \pi^+$ 
in IRA without considering the broken $SU(3)_f$ symmetry~\cite{Geng:2017mxn,Geng:2019xbo,Geng:2017esc,Geng:2018plk,Jia:2019zxi}.
Since ${\cal R}(\Xi_c^0)$ suggests the existence of the broken effect, 
we add $E_{\bf B}^s=E_M^{\prime\,s}=|E_{\bf B}^s|e^{i\delta_{E_{\bf B}^s}}$
to the parameters in Eq.~(\ref{para}) for the S2 global fit.
We thus determine the parameters in the two scenarios (S1 and S2), given in Table~\ref{fit_result}.
Moreover, we present our calculations in Tables~\ref{BF_CA}, \ref{BF_SCS} and \ref{BF_DCS}
for ${\cal B}({\bf B}_c\to{\bf B}M)$ to be compared with the experimental results 
and other theoretical calculations.

%
\begin{table}[t!]
\caption{Doubly Cabibbo-suppressed (DCS) amplitudes
in the expansions of ${\cal M}_{\text{TDA}}$ and ${\cal M}_{\text{IRA}}$.}\label{tab3}
\tiny
\begin{tabular}{|l|l|l|}
\hline
Decay mode
&$\;\;\;\;\;\;\;\;\;\;\;\;\;\;\;\;\;\;\;\;\;\;\;\;$ $M_{\text{TDA}}$
&$\;\;\;\;\;\;\;\;\;\;\;\;\;\;\;\;\;\;\;\;\;\;\;\;$ $ M_{\text{IRA}}$ \\
\hline\hline
$\Lambda_c^+ \to p K^0$
&$(C^\prime-2C)s_c^2$
& $ -2(a_{3}-\frac{a_{4} + a_{6}}{2})s_c^2$
\\
$\Lambda_c^+ \to n  K^+$
&$-(C^\prime +2T)s_c^2$
&$2(a_{3}+\frac{a_{4} + a_{6}}{2})s_c^2$
\\
\hline
$\Xi_c^+ \to  \Lambda^0 K^+$
&$\frac{1}{\sqrt{6}}(2C+E_{{\bf B}}^{(s)}-2E^{\prime(s)} +2T)s_c^2$
& $\frac{\sqrt{6}}{3}(a_1-2a_2-2a_3-\frac{a_5+a_6-2a_7}{2})s_c^2$
\\
$\Xi_c^+ \to  \Sigma^0 K^+$
&$\frac{1}{\sqrt{2}}(E_{{\bf B}}^{(s)}-2T)s_c^2$
&$ \sqrt 2(a_{1}-\frac{a_{5} - a_{6}}{2})s_c^2$
\\
$\Xi_c^+ \to \Sigma^+ K^0$
&$(2C-E_M^{\prime(s)})s_c^2$
&$ 2(a_{1}-\frac{a_{5} + a_{6}}{2})s_c^2$
\\
$\Xi_c^+ \to n\pi^+$
&$-E^\prime s_c^2$
& $2(a_{2}-\frac{a_{4} + a_{7}}{2})s_c^2$
\\
$\Xi_c^+ \to p \pi^0$
&$\frac{1}{\sqrt 2}(E^\prime-E_{\bf B}-E_M^\prime)s_c^2$
&$\sqrt 2(a_{2}+\frac{a_{4} - a_{7}}{2})s_c^2$
\\
$\Xi_c^+ \to p \eta $
&$[\frac{1}{\sqrt 2}(E^\prime+E_{\bf B}-E_M^\prime) c\phi+C^\prime s\phi]s_c^2 $
& $-[\sqrt{2}c\phi(a_2-\frac{a_4+a_7}{2})-2s\phi(a_1 - a_3- \frac{a_5 }{2})]s_c^2$
\\
%
\hline
$\Xi_c^0 \to \Lambda^0 K^0$
&$-\frac{1}{\sqrt{6}}(2C-2C^\prime+2E_{M}^{(s)}+E_M^{\prime(s)}+2E^{\prime(s)})s_c^2$
&$\frac{\sqrt{6}}{3}(a_1-2a_2-2a_3+\frac{a_5+a_6-2a_7}{2})s_c^2 $
\\
$\Xi_c^0 \to \Sigma^0 K^0$
&$\frac{1}{\sqrt{2}}(-2C+E_M^{\prime(s)})s_c^2$
&$ -\sqrt 2(a_{1}+\frac{a_{5} - a_{6}}{2})s_c^2$
\\
$\Xi_c^0 \to \Sigma^- K^+$
&$(-2T+E_{\bf B}^{(s)})s_c^2$
& $2(a_{1}+\frac{a_{5} + a_{6}}{2})s_c^2$
\\
$\Xi_c^0 \to n \pi^0$
&$-\frac{1}{\sqrt 2}(E_{{\bf B}}+E_M^\prime+E_{M}+E^\prime)s_c^2$
&$-\sqrt 2(a_{2}-\frac{a_{4} - a_{7}}{2})s_c^2$
\\
$\Xi_c^0 \to p \pi^-$
&$(E_{M}+E^\prime)s_c^2$
&$2(a_{2}+\frac{a_{4} + a_{7}}{2})s_c^2$
\\
$\Xi_c^0 \to n \eta $
&$[\frac{1}{\sqrt 2}(E_{\bf B}-E_M^\prime+E_{M}+E^\prime)c\phi+C^\prime s\phi ]s_c^2$
&$[\sqrt{2}c\phi( a_2 + \frac{a_4 +a_7}{2} )-2s\phi( a_1 -a_3 +\frac{a_5}{2})]s_c^2$
\\
%
\hline
\end{tabular}
\end{table}
%
%
\begin{table}[b!]
\caption{Fit results of the topological parameters in S1 and S2 
for the exact and broken $SU(3)_f$ symmetries, respectively.
Besides, $(|T|,|C^{(\prime)}|,|E_{\bf B}^{(s)}|,|E_M|,|E'|)$ are in units of GeV$^3$,
and $n.d.f$ the number of degrees of freedom.}\label{fit_result}
\scriptsize
\begin{tabular}{|c|cc|}
\hline
&TDA (S1)& TDA (S2)
\\
\hline\hline
$\chi^2$&4.5&5.7\\
$n.d.f$&5&4\\
$|T|$&$0.23\pm0.02$&$0.24\pm0.02$\\
$|C|$&$0.26\pm0.01$&$0.23\pm0.02$\\
$|C^\prime|$&$0.34\pm0.02$&$0.32\pm0.03$\\
$|E_{\bf B}|$&$0.22\pm0.03$&$0.22\pm0.05$\\
$|E_{\bf B}^s|$&&$0.37\pm0.06$\\
$|E_M|$&$0.40\pm0.03$&$0.38\pm0.03$\\
$|E^\prime|$&$0.24\pm0.02$&$0.23\pm0.02$\\
$\delta_C$&$(183.2\pm9.6)^\circ$&$(179.5\pm12.9)^\circ$\\
$\delta_{C^\prime}$&$(163.7\pm5.0)^\circ$&$(149.7\pm6.7)^\circ$\\
$\delta_{E_{\bf B}}$&$(-100.3\pm7.1)^\circ$&$(-93.6\pm8.2)^\circ$\\
$\delta_{E_{\bf B}^s}$&&$(43.3\pm8.0)^\circ$\\
$\delta_{E_M}$&$(100.3\pm8.0)^\circ$&$(113.2\pm10.7)^\circ$\\
$\delta_{E^\prime}$&$(-71.1\pm6.7)^\circ$&$(-50.1\pm12.3)^\circ$\\
\hline
\end{tabular}
\end{table}
%
%
%
\begin{table}[t!]
\caption{Cabibbo-allowed (CA) branching fractions.}\label{BF_CA}
{
\tiny
\begin{tabular}{|l|ccccc|c|}
\hline
$\,$ Branching fraction
&pole model~\cite{Cheng:2018hwl,Zou:2019kzq}
&IRA~\cite{Geng:2019xbo,IRA1c}
&IRA~\cite{Huang:2021aqu}
&TDA~\cite{Zhao:2018mov}
&TDA [This work] (S1, S2)&Data
\\
\hline\hline
$10^2{\cal B}(\Lambda_c^+\to{\Lambda^{0} \pi^{+}})$
&1.30&$1.27\pm0.07$&$1.307\pm0.069$&$1.32\pm0.34$
&$(1.27 \pm0.22,1.24\pm 0.30)$
&$1.31\pm0.09$~\cite{BESIII:2022bkj}\\
$10^2{\cal B}(\Lambda_c^+\to{\Sigma^{0} \pi^{+}})$
&2.24&$1.26 \pm 0.06$&$1.272\pm0.056$&$1.26\pm0.32$
&$(1.22 \pm 0.23,1.24 \pm 0.30)$
&$1.22\pm0.11$~\cite{BESIII:2022bkj}\\
$10^2{\cal B}(\Lambda_c^+\to{\Sigma^{+} \pi^{0}})$
&2.24&$1.26 \pm 0.06$&$1.283\pm0.057$&$1.23\pm0.17$
&$(1.22\pm 0.23,1.24 \pm 0.30)$
&$1.25\pm0.10$~\cite{pdg}\\
$10^2{\cal B}(\Lambda_c^+\to{\Xi^{0} K^{+}})$
&0.73&$0.57\pm0.09$&$0.548\pm0.068$&$0.59\pm0.17$
&$(0.54 \pm0.07,0.51 \pm 0.07)$
&$0.55\pm0.07$~\cite{pdg}\\
$10^2{\cal B}(\Lambda_c^+\to{p \bar{K}^{0}})$
&2.11&$3.14 \pm 0.15$&$3.174\pm0.154$&$3.14\pm1.00$
&$(3.18 \pm 0.64,3.10 \pm 0.80)$
&$3.18\pm0.16$~\cite{pdg}\\
$10^2{\cal B}(\Lambda_c^+\to{\Sigma^{+} \eta})$
&0.74&$0.29\pm0.12$&$0.45\pm0.19$&$0.47\pm0.22$
&$(0.42\pm0.18,0.57 \pm0.26)$
&$0.44\pm0.20$~\cite{pdg}\\
\hline
$10^3{\cal B}(\Xi_c^+\to{\Sigma^{+} \bar{K}^{0}})$
&2.0&$7.8^{+10.2}_{-\;\,7.8}$&$10.6\pm14.0$&$24.1\pm7.1$
&$(12.7\pm 7.0,14.7^{+10.7}_{-\,\,8.4})$
&\\
$10^3{\cal B}(\Xi_c^+\to{\Xi^{0} \pi^{+}})$
&17.2&$4.2\pm1.7$&$5.4\pm1.8$&$9.3\pm3.6$
&$(7.0^{+3.3}_{-2.6},15.7\pm 6.2)$
&$16.0\pm 8.0$~\cite{pdg}\\
\hline
$10^3{\cal B}(\Xi_c^0\to{\Lambda^{0} \bar{K}^{0}})$
&13.3&$14.2 \pm 0.09$
&$6.68\pm1.30$&$8.3\pm5.0$
&$(9.85\pm 2.26,10.0\pm 2.9)$
&$8.24\pm2.44$~\cite{Belle:2021avh}\\
$10^3{\cal B}(\Xi_c^0\to{\Sigma^{0} \bar{K}^{0}})$
&$0.4$&$0.9^{+1.1}_{-0.9}$
&$1.38\pm0.48$&$7.9\pm4.8$
&$(1.48^{+1.27}_{-0.92},1.46^{+1.57}_{-1.09})$
&$1.38\pm0.48$~\cite{Belle:2021avh}\\
$10^3{\cal B}(\Xi_c^0\to{\Sigma^{+} K^{-}})$
&$4.6$&$7.6 \pm 1.4$
&$2.21\pm0.68$&$22.0\pm5.7$
&$(2.21^{+0.37}_{-0.16},2.25^{+1.8}_{-1.1})$
&$2.21\pm0.68$~\cite{Belle:2021avh}\\
$10^3{\cal B}(\Xi_c^0\to{\Xi^{0} \pi^{0}})$
&$18.2$&$10.0 \pm 1.4$
&$2.56\pm0.93$&$4.7\pm0.9$
&$(6.0\pm1.2,3.6\pm1.2)$
&\\
$10^3{\cal B}(\Xi_c^0\to{\Xi^{-} \pi^{+}})$
&64.7&$29.5 \pm 1.4$
&$12.1\pm2.1$&$19.3\pm2.8$
&$(24.5\pm 3.7,23.3\pm 4.5)$
&$18.0\pm 5.2$~\cite{Belle:2021avh}\\
$10^3{\cal B}(\Xi_c^0\to{\Xi^{0} \eta})$
&26.7&$13.0\pm 2.3$
&&$8.3\pm2.3$
&$(4.2^{+1.6}_{-1.3},7.3\pm3.2)$
&\\
\hline
\end{tabular}}
\end{table}
%
\section{Discussions and Conclusions}
Under the group theoretical consideration, we present the seven $SU(3)_f$ singlets for ${\bf B}_c\to{\bf B}M$,
which are in agreement with the seven independent IRA amplitudes in Eq.~(\ref{amp1}).
Since TDA also relies on the $SU(3)_f$ symmetry, there should exist seven independent TDA amplitudes. 
We draw and parameterize eight TDA amplitudes as
those using the topological-diagram scheme~\cite{Kohara:1991ug,Chau:1995gk}.
By finding that either $E'$ or $E_{\bf B}^\prime$ is redundant, we reduce them to seven.
We hence obtain the unique relations in Eqs.~(\ref{relation1}) and (\ref{relation2}),
and demonstrate that TDA and IRA are the equivalent $SU(3)_f$ approaches.

Confusingly, there can be six, seven and sixteen topological amplitudes
from Refs.~\cite{Zhao:2018mov},~\cite{Groote:2021pxt} and~\cite{He:2018joe}, respectively.
In Ref.~\cite{Zhao:2018mov}, the less TDA amplitudes reflects the fact that
one disregards the quark orderings for ${\bf B}_{(c)}$. It also fails to present the isospin relation:
${\cal M}(\Lambda_c^+ \to \Sigma^0 \pi^+)=-{\cal M}(\Lambda_c^+ \to \Sigma^+ \pi^0)$.
Although Ref.~\cite{Groote:2021pxt} provides the seven TDA parameters, the equality of
${\cal M}(\Xi_c^0\to \Xi^- K^+)=-s_c{\cal M}(\Xi_c^0\to \Xi^- \pi^+)$ is not given, 
which disagrees with the $SU(3)_f$ approaches. 
In Ref.~\cite{He:2018joe}, 
the sixteen amplitudes are due to that all possible quark orderings of the octet baryon are taken into account,
which can be reduced with the identity 
${\bf B}_{ijk} + {\bf B}_{kij} + {\bf B}_{jki} = 0$~\cite{Kohara:1991ug,Pan:2020qqo}.
%
\begin{table}[t!]
\caption{Singly Cabibbo-suppressed (SCS) branching fractions.}\label{BF_SCS}
{
\tiny
\begin{tabular}{|l|ccccc|c|}
\hline
$\,$ Branching fraction
&pole model~\cite{Cheng:2018hwl,Zou:2019kzq}
&IRA~\cite{Geng:2019xbo,IRA1c} 
&IRA~\cite{Huang:2021aqu}
&TDA~\cite{Zhao:2018mov}
&TDA [This work] (S1, S2)&Data
\\
\hline\hline
$10^4{\cal B}(\Lambda_c^+\to{\Lambda^{0} K^{+}})$
&10.7&$6.6\pm0.9$&$6.4\pm1.0$&$5.9\pm1.7$
&$(6.6 \pm1.4,6.9 \pm 1.6)$
&$6.1\pm1.2$~\cite{pdg}\\
$10^4{\cal B}(\Lambda_c^+\to{\Sigma^{0} K^{+}})$
&7.2&$5.2 \pm 0.7$&$5.04\pm0.56$&$5.5\pm1.6$
&$(4.2 \pm0.8,4.9 \pm 1.6)$
&$5.2\pm0.8$~\cite{pdg}\\
$10^4{\cal B}(\Lambda_c^+\to{\Sigma^{+} K^{0}})$
&14.4&$10.5\pm 1.4$&$2.06\pm0.84$&$19.1\pm4.8$
&$(10.1\pm 1.8,17.5\pm 4.2)$
&\\
$10^4{\cal B}(\Lambda_c^+\to{n \pi^{+}})$
&2.7&$7.6\pm 1.1$&$3.5\pm1.1$&$7.7\pm2.0$
&$(7.6\pm 1.7,8.3 \pm2.6)$
&$6.6\pm1.3$~\cite{BESIII:2022bkj}\\
$10^4{\cal B}(\Lambda_c^+\to{p \pi^{0}})$
&1.3&$1.1^{+1.3}_{-1.1}$&$44.5\pm8.5$&$0.8^{+0.9}_{-0.8}$
&$(0.3^{+1.0}_{-0.3},0.4^{+1.7}_{-0.4})$
&$0.3\pm0.3\,(<0.8)$~\cite{Belle:2021mvw}\\
$10^4{\cal B}(\Lambda_c^+\to{p \eta})$
&12.8&$11.2\pm2.8$&$12.7\pm2.4$&$11.4\pm3.5$
&$(14.2\pm2.3,14.7\pm 2.8)$
&$14.2\pm1.2$~\cite{Belle:2021mvw}\\
\hline
$10^4{\cal B}(\Xi_c^+\to{\Lambda^{0} \pi^{+}})$
&8.5&$12.3 \pm 4.2$&$5.6\pm1.4$&$13.9\pm5.1$
&$(2.1\pm 1.1,0.2^{+0.5}_{-0.4})$
&\\
$10^4{\cal B}(\Xi_c^+\to{\Sigma^{0} \pi^{+}})$
&43.0&$26.5 \pm 2.5$&$31.8\pm2.7$&$13.4\pm4.9$
&$(13.4\pm 2.0,13.3\pm 2.4)$
&\\
$10^4{\cal B}(\Xi_c^+\to{\Sigma^{+} \pi^{0}})$
&13.6&$26.1 \pm 6.7$&$24.0\pm20.0$&$16.4\pm3.2$
&$(28.2 \pm 4.1,28.1\pm 5.6)$
&\\
$10^4{\cal B}(\Xi_c^+\to{\Xi^{0} K^{+}})$
&22.0&$7.6 \pm 1.6$&$9.0\pm12.0$&$12.3\pm3.1$
&$(7.7\pm 2.5,14.1\pm 5.1)$
&\\
$10^4{\cal B}(\Xi_c^+\to{p \bar{K}^{0}})$
&39.6&$46.4 \pm 7.2$&$19.8\pm16.6$&$48.6\pm12.2$
&$(25.3 \pm 4.6,30.7\pm 7.5)$
&\\
$10^4{\cal B}(\Xi_c^+\to{\Sigma^{+} \eta})$
&3.2&$15.0\pm 10.6$&$12.0\pm13.0$&$14.1\pm3.9$
&$(9.1\pm 3.2,24.0\pm 8.2)$
&\\
\hline
$10^4{\cal B}(\Xi_c^0\to{\Sigma^{+} \pi^{-}})$
&7.1&$4.9 \pm 0.9$
&$1.31\pm0.49$&$2.4\pm1.5$
&$(1.3^{+0.6}_{-0.5},1.3^{+1.1}_{-0.6})$
&\\
$10^4{\cal B}(\Xi_c^0\to{\Sigma^{-} \pi^{+}})$
&26.2&$18.3 \pm 0.9$
&$8.0\pm1.4$&$11.1\pm1.6$
&$(9.0 \pm 2.5,8.5\pm 3.9)$
&\\
$10^4{\cal B}(\Xi_c^0\to{\Lambda^{0} \pi^{0}})$
&2.4&$3.1\pm 1.1$&$4.4\pm1.2$&$1.0\pm0.5$
&$(0.8\pm 0.5,0.9\pm 0.6)$
&\\
$10^4{\cal B}(\Xi_c^0\to{\Sigma^{0} \pi^{0}})$
&3.8&$5.0 \pm 0.9$
&$0.32\pm0.90$&$1.0\pm0.5$
&$(0.4 \pm 0.1,0.3\pm 0.1)$
&\\
$10^4{\cal B}(\Xi_c^0\to{\Xi^{-} K^{+}})$
&39.0&$12.8 \pm 0.6$&$4.70\pm0.83$&$5.6\pm0.8$
&$(11.8\pm 1.8,4.1 \pm 2.8)$
&$3.9\pm1.2$~\cite{pdg}\\
$10^4{\cal B}(\Xi_c^0\to{\Xi^{0} K^{0}})$
&13.2&$9.6 \pm 0.4$&$3.8\pm2.6$&$6.3\pm1.9$
&$(4.2^{+1.8}_{-1.5},14.6\pm5.4)$
&\\
$10^4{\cal B}(\Xi_c^0\to{p K^{-}})$
&3.5&$6.0 \pm 1.3$&$2.6\pm7.8$&$2.5\pm1.6$
&$(1.3^{+0.6}_{-0.5},1.3^{+1.1}_{-0.6})$
&\\
$10^4{\cal B}(\Xi_c^0\to{n \bar{K}^{0}})$
&14.0&$10.7 \pm 0.6$&$3.6\pm6.6$&$7.8\pm2.3$
&$(5.0^{+2.2}_{-1.8},5.6\pm2.5)$
&\\
$10^4{\cal B}(\Xi_c^0\to{\Lambda^{0} \eta})$
&8.1&$7.9\pm2.7$&&$2.6\pm1.3$
&$(2.2\pm0.6,0.6\pm 0.4)$
&\\
$10^4{\cal B}(\Xi_c^0\to{\Sigma^{0} \eta})$
&0.5&$1.8\pm1.1$&&$2.5\pm1.3$
&$(4.0\pm1.6,6.1\pm 1.7)$
&\\
\hline
\end{tabular}}
\end{table}
%

The equivalent $SU(3)_f$ approaches are able to provide more information 
on the hadronization in the ${\bf B}_c\to{\bf B}M$ decays.
For instance, we derive $E_M^\prime=E_{\bf B}=a_4$ 
to reduce the topological diagrams involved in the decays,
and $a_1=(T-C)/2$ indicates that $a_1$ connects the two $W$-emission topological amplitudes.
With $a_5=0$, $a_5$ has no topological correspondence. 
The QCD-disfavored parameters have been commonly neglected in the numerical analyses~\cite{Lu:2016ogy,Geng:2017mxn,Geng:2019awr,Geng:2019xbo,
Geng:2017esc,Geng:2018plk,Geng:2018upx,Hsiao:2019yur,Geng:2020zgr},
such as $(a_4,a_5,a_6,a_7)$ in the ${\bf B}_c\to{\bf B}M$ decays. 
According to Eq.~(\ref{relation2}),
we find that $(a_4,a_6,a_7)$ are associated with the topological parameters
which can be sizeable. Therefore, it is unlikely that the QCD disfavored parameters are negligible, 
whereas they were commonly discarded.

In Eq.~(\ref{RXic}), since ${\cal R}(\Xi_c^0)$ has indicated that 
one cannot explain ${\cal B}_{ex}(\Xi_c^0\to\Xi^{-} K^{+})$ with the exact $SU(3)_f$ symmetry,
it has been excluded in the S1 global fit. As a result, $\chi^2/n.d.f=0.9$ presents a reasonable fit.
We also include ${\cal B}_{ex}(\Xi_c^0\to\Xi^{-} K^{+})$ for a test, which causes $\chi^2/n.d.f=5.5$.
We hence add $E_{\bf B}^s=E_M^{\prime\,s}=|E_{\bf B}^s|e^{i\delta_{E_{\bf B}^s}}$ 
as the new parameters in the S2 global fit, in order to accommodate ${\cal B}_{ex}(\Xi_c^0\to\Xi^{-} K^{+})$.
It turns out to be a reasonable fit with $\chi^2/n.d.f=1.4$, and 
${\cal B}_{th}(\Xi_c^0\to\Xi^{-} K^{+})=(4.1 \pm 2.8)\times 10^{-4}$ can explain the data.
Moreover, we extract $|n_q|=1.2\pm 0.4$ and $\delta_{n_q}=(50.3\pm 11.5)^\circ$ 
in $E_{\bf B}^s=|n_q|e^{i\delta_{n_q}} E_{\bf B}$ as the measure of the $SU(3)_f$ symmetry breaking.
One can study the broken $SU(3)_f$ symmetry with IRA~\cite{Savage:1991wu,Geng:2018bow}.
Without the topological indication in Eq.~(\ref{RXic}), three IRA amplitudes should be introduced
for the broken effects in ${\bf B}_c\to{\bf B}M$~\cite{Geng:2018bow}.

%
\begin{table}[t!]
\caption{Doubly Cabibbo-suppressed (DCS) branching fractions.}\label{BF_DCS}
{
\tiny
\begin{tabular}{|l|ccccc|c|}
\hline
$\,$ Branching fraction
&pole model~\cite{Cheng:2018hwl,Zou:2019kzq}
&IRA~\cite{Geng:2019xbo,IRA1c} 
&IRA~\cite{Huang:2021aqu,IRA2c}
&TDA~\cite{Zhao:2018mov}
&TDA [This work] (S1, S2)&Data
\\
\hline\hline
$10^5{\cal B}(\Lambda_c^+\to{p K^{0}})$
&&$1.2^{+1.4}_{-1.2}$&$1.5\pm 5.6$
&$3.7\pm1.1$
&$(1.8^{+1.0}_{-0.8},2.1^{+1.5}_{-1.2})$
&\\
$10^5{\cal B}(\Lambda_c^+\to{n K^{+}})$
&&$0.4\pm0.2$&$4.8\pm2.2$&$1.4\pm0.5$
&$(1.0 \pm0.5,2.2\pm 0.9)$&\\
\hline
$10^5{\cal B}(\Xi_c^+\to{\Lambda^{0} K^{+}})$
&&$3.3 \pm 0.8$&$3.65\pm0.53$&$7.5\pm1.9$
&$(1.3^{+0.7}_{-0.6},4.6\pm 2.1)$
&\\
$10^5{\cal B}(\Xi_c^+\to{\Sigma^{0} K^{+}})$
&&$11.9 \pm 0.7$&$12.23\pm0.57$&$7.2\pm1.8$
&$(10.3\pm 1.6,3.6\pm 1.4)$
&\\
$10^5{\cal B}(\Xi_c^+\to{\Sigma^{+} K^{0}})$
&&$19.5 \pm 1.7$&$34.6\pm 2.2$ 
&$16.9\pm5.4$
&$(18.1 \pm 3.6,41.9\pm 8.1)$
&\\
$10^5{\cal B}(\Xi_c^+\to{n \pi^{+}})$
&&$12.1 \pm 2.8$&$13.2\pm3.5$&$5.2\pm1.5$
&$(4.7 \pm 0.6,4.5\pm 0.6)$
&\\
$10^5{\cal B}(\Xi_c^+\to{p \pi^{0}})$
&&$6.0 \pm 1.4$&$40.0\pm49.0$&$1.5\pm1.5$
&$(2.8^{+1.6}_{-1.3},4.2^{+3.2}_{-2.4})$
&\\
$10^5{\cal B}(\Xi_c^+\to{p \eta})$
&&$20.4\pm8.4$&$50.0\pm120.0$&$16.6\pm3.1$
&$(7.0\pm0.8,7.9\pm 1.0)$
&\\
\hline
$10^5{\cal B}(\Xi_c^0\to{\Lambda^{0} K^{0}})$
&&$0.6 \pm 0.2$&$0.37\pm 0.84$ 
&$2.4\pm1.4$
&$(0.1^{+0.2}_{-0.1},0.8^{+0.8}_{-0.6})$
&\\
$10^5{\cal B}(\Xi_c^0\to{\Sigma^{0} K^{0}})$
&&$2.5 \pm 0.2$&$4.12\pm 0.19$ 
&$2.3\pm1.4$
&$(3.0 \pm 0.9,7.0\pm 1.4)$
&\\
$10^5{\cal B}(\Xi_c^0\to{\Sigma^{-} K^{+}})$
&&$6.1 \pm 0.4$&$3.28\pm0.58$&$5.5\pm0.7$
&$(6.9\pm 1.0,2.4 \pm 0.9)$
&\\
$10^5{\cal B}(\Xi_c^0\to{n \pi^{0}})$
&&$1.5 \pm 0.4$&$2.6\pm2.7$&$3.3\pm0.9$
&$(1.1^{+0.7}_{-0.5},1.1^{+1.1}_{-0.8})$
&\\
$10^5{\cal B}(\Xi_c^0\to{p \pi^{-}})$
&&$3.1 \pm 0.7$&$1.4\pm43.0$&$7.6\pm2.0$
&$(0.7^{+0.4}_{-0.3},0.8\pm0.6)$
&\\
$10^5{\cal B}(\Xi_c^0\to{n \eta})$
&&$5.2\pm 2.1$&&$4.2\pm0.8$
&$(1.1^{+0.4}_{-0.2},0.8^{+0.4}_{-0.3})$
&\\
\hline
\end{tabular}}\end{table}
%

As can be seen in Tables~\ref{BF_CA} and \ref{BF_SCS}, 
the pole model~\cite{Cheng:2018hwl,Zou:2019kzq}, 
IRA with the neglecting of $(a_4,a_5,a_7)$~\cite{Geng:2019xbo},
and TDA of this work (S1) all lead to ${\cal R}(\Xi_c^0)\simeq 0.05$.
It seems that TDA of Ref.~\cite{Zhao:2018mov} gives the consistent ${\cal R}(\Xi_c^0)\simeq 0.03$;
it is, however, based on ${\cal M}(\Xi_c^0\to \Xi^- K^+)\ne -s_c{\cal M}(\Xi_c^0\to \Xi^- \pi^+)$
inconsistent with the other $SU(3)_f$ approaches.
By contrast, the recent global fit using IRA without neglecting the QCD-disfavored parameters
can present ${\cal R}(\Xi_c^0)\simeq 0.04$ close to the data~\cite{Huang:2021aqu},
where the amplitudes are considered to depend on the actual particle masses.

According to our numerical results, the equivalent $SU(3)_f$ approaches 
are able to interpret the new data, that is, 
${\cal B}(\Xi_c^0\to \Lambda^0 \bar K^0,\Sigma^0 \bar K^0,\Sigma^+ K^-)$ and
${\cal B}(\Lambda_c^+ \to n \pi^+,p\pi^0,p\eta)$. Particularly, we find that
the topology $E_M$ only contributes to $\Xi_c^0\to{\bf B}M$,
but neglected in the pole model of Ref.~\cite{Zou:2019kzq} 
and IRA of Ref.~\cite{Geng:2019xbo}. $E_M$ plays a key role in
explaining ${\cal B}(\Xi_c^0\to \Lambda^0 \bar K^0,\Sigma^0 \bar K^0,\Sigma^+ K^-)$.
For example, since $E_M=-2a_4-2a_7$ largely cancels
$E'=-2a_2+a_4+a_7$ in ${\cal M}(\Xi_c^0 \to \Sigma^+ K^-)=(E_{M}+E^\prime)s_c$,
${\cal B}(\Xi_c^0 \to \Sigma^+ K^-)$ can be as small as $2.0\times 10^{-3}$.
As a test, we set $E_M=0$, which results in 
${\cal B}(\Xi_c^0\to \Lambda^0 \bar K^0,\Sigma^0 \bar K^0,\Sigma^+ K^-)
\simeq (19.3,4.6,4.4)\times 10^{-3}$ deviating from the data.

For $\Lambda_c^+ \to n \pi^+$ and $\Lambda_c^+ \to p\pi^0$, 
the pole model predicts the destructive interferences between the factorizable and non-factorizable amplitudes,
leading to ${\cal B}_{th}(\Lambda_c^+\to n \pi^{+},p \pi^{0})\simeq (2.7,1.3)\times 10^{-4}$~\cite{Cheng:2018hwl},
whereas ${\cal B}_{ex}(\Lambda_c^+\to n \pi^{+})
\gg{\cal B}_{ex}(\Lambda_c^+\to p\pi^0)$~\cite{BESIII:2022bkj,Belle:2021mvw}.
Here, we can present ${\cal M}(\Lambda_c^+\to n \pi^{+},p\pi^0)\sim (A+B,A-B)$
with $(A,B)=(a_2+a_3-a_7/2,a_6/2)$, such that $\Lambda_c^+\to n \pi^{+}$ and $\Lambda_c^+\to p\pi^0$
can be viewed to proceed through the constructive and destructive interferences, respectively.
Since the $SU(3)_f$ symmetry predicts ${\cal M}(\Xi_c^+\to \Xi^0 K^+)={\cal M}(\Lambda_c^+\to n \pi^{+})$,
one should have ${\cal B}(\Xi_c^+\to \Xi^0 K^+)\simeq
{\cal R}_d{\cal B}_{ex}(\Lambda_c^+\to n \pi^{+})\simeq (12-18)\times 10^{-4}$
with ${\cal R}_d=2.4$ the ratio of the dynamical factors in Eq.~(\ref{p_space}),
which is consistent with our predictions in Table~\ref{BF_SCS}.
Therefore, ${\cal B}(\Xi_c^+\to \Xi^0 K^+)$ compared to future measurements
can be used to test the validity of the $SU(3)_f$ approaches.

\newpage
In summary, 
we have studied the two-body anti-triplet charmed baryon decays using
the irreducible $SU(3)_f$ approach (IRA) and topological-diagram approach (TDA).
Due to the group theoretical consideration, 
we have presented that there can be the same number of the IRA and TDA amplitudes.
We have hence found out the unique relations, and demonstrated that 
IRA and TDA are the equivalent $SU(3)_f$ approaches. We have explained
the recently measured branching fractions, that is,
${\cal B}(\Xi_c^0\to \Lambda^0 \bar K^0,\Sigma^0 \bar K^0,\Sigma^+ K^-)$ and
${\cal B}(\Lambda_c^+ \to n \pi^+,p\pi^0,p\eta)$. 
Moreover, we have predicted the branching fractions of ${\bf B}_c\to{\bf B}M$ 
under the exact and broken $SU(3)_f$ symmetries, 
which can be tested by future measurements.

\section*{ACKNOWLEDGMENTS}
We would like to thank Prof.~X.~G. He, Prof.~C.~P. Shen and Dr.~Yang Li
for useful discussions. We would also like to thank Prof.~Wei Wang for valuable comments,
and Dr.~Jin~Sun for pointing out the typos. This work was supported in part
by National Science Foundation of China (Grants No.~11675030 and No.~12175128).


\end{document}